\documentclass[12pt]{article}

\usepackage[utf8]{inputenc}
\usepackage{amsmath, amssymb}
\usepackage{graphicx}
\usepackage{hyperref}
\usepackage{listings}
\usepackage{xcolor}
\usepackage{authblk}  
\usepackage[numbers]{natbib}
\usepackage{fancyhdr}  

\title{nsEVDx: A Python library for modeling Non-Stationary Extreme Value Distributions}

\author[1]{Nischal Kafle\thanks{Corresponding author: \href{mailto:nkafle.29@gmail.com}{nkafle.29@gmail.com}. ORCID: 0009-0004-3187-4920}}
\author[1]{Claudio Meier}

\affil[1]{Department of Civil Engineering, University of Memphis, TN, USA}

\date{2025-09-08}

\begin{document}

\maketitle

\section{Summary}

\texttt{nsEVDx} is an open-source Python package for fitting stationary and non-stationary Extreme Value Distributions (EVDs) to extreme value data. It can be used to model extreme events in fields like hydrology, climate science, finance, and insurance, using both frequentist and Bayesian methods. For Bayesian inference it employs advanced Monte Carlo sampling techniques such as Metropolis-Hastings, Metropolis-adjusted Langevin (MALA), and Hamiltonian Monte Carlo (HMC). Unlike many existing extreme value theory (EVT) tools, which can be complex or lack Bayesian options, \texttt{nsEVDx} offers an intuitive, Python-native interface that is both user-friendly and extensible. It requires only standard scientific Python libraries (\texttt{numpy}, \texttt{scipy}) for its core functionality, while optional features like plotting and diagnostics use \texttt{matplotlib} and \texttt{seaborn}. A key feature of \texttt{nsEVDx} is its flexible support for non-stationary modeling, where the location, scale, and shape parameters can each depend on arbitrary, user-defined covariates. This enables practical applications such as linking extremes to other variables (e.g., rainfall extremes to temperature or maximum stock market losses to market volatility indices). Overall, \texttt{nsEVDx} aims to serve as a practical, easy-to-use, and extensible tool for researchers and practitioners analyzing extreme events in non-stationary environments.

\section{Statement of Need}

Probabilistic modeling of extreme events is essential across disciplines, from resilient infrastructure design and climate adaptation to insurance pricing and financial risk management. In many real-world processes, the statistical properties of the extremes are often non-stationary, driven by long-term changes such as climate change, urbanization, or economic shifts. Accurately estimating return periods and risks under these evolving conditions requires fitting non-stationary extreme value distributions (EVDs) to observations.

Several R packages currently support EVD modeling, including \texttt{ismev} \cite{heffernan_j_e_ismev_2003}, \texttt{extRemes} \cite{gilleland_extremes_2025}, \texttt{climextRemes} \cite{paciorek_climextremes_2016}, and \texttt{NSGEV} \cite{irsn_nsgev_2024}. However, these packages differ in their ability to handle non-stationary models and Bayesian inference. Moreover, extending their functionality and integrating modern inference techniques can be challenging. Probabilistic programming frameworks, such as python-based \texttt{PyMC} \cite{oriol_abril-pla_pymc_2023}, and C++ based Stan with interfaces like \texttt{PyStan} \cite{noauthor_pystan_2023} and \texttt{CmdStan} \cite{noauthor_cmdstan_2023}, offer powerful tools for building custom statistical models, including those for extreme value analyses. However, these tools require significant expertise in both statistics and programming to develop, tune, and validate the models effectively. As a result, they may be too complex for domain experts like hydrologists, climate scientists, or risk analysts seeking easy-to-use methods.

Based on this synopsis, there is a clear need for a Python tool that balances flexibility and ease of use, while supporting arbitrary covariates, parameter constraints, custom priors, and advanced MCMC algorithms such as MALA \cite{roberts_exponential_1996} and HMC \cite{michael_betancourt_conceptual_2017}, for fitting non-stationary Generalized Extreme Value (GEV) and Generalized Pareto (GPD) distributions, the two most prominent EVDs.

To bridge this gap, we developed \texttt{nsEVDx}, a flexible, user-friendly Python package that streamlines non-stationary EVD modeling without compromising statistical rigor. Developed as part of N. Kafle's PhD research, \texttt{nsEVDx} has been applied in hydrology \cite{kafle_evaluating_2025} and is applicable to fields like climate science, finance, and engineering, where it is critical to understand the frequency and intensity of extremes under non-stationarity conditions. Its application is also reflected in an upcoming technical paper on trends in short-duration extreme rainfall in the Southeastern U.S. \cite{kafle_detecting_nodate}.

\section{Features}

\begin{itemize}
    \item Supports both the Generalized Extreme Value (GEV) and Generalized Pareto (GPD) distributions
    \item Non-stationary modeling via linear and log-linear relationships between parameters and covariates
    \item Independent non-stationarity in location, scale, and shape parameters
    \item Frequentist and Bayesian inference support
    \item MCMC algorithms: Random Walk, Metropolis-adjusted Langevin (MALA), and Hamiltonian Monte Carlo (HMC)
    \item Custom priors, parameter bounds, and temperature scaling for tuning MCMC
    \item Integrated diagnostics: trace plots, convergence checks, and posterior visualization
    \item Modular and extensible API designed for ease of use by domain scientists
    \item Bayesian metrics and likelihood ratio tests
\end{itemize}

\section{Implementation}

The core class \texttt{NonStationaryEVD} handles parameter parsing, log-likelihood construction, prior specification, and proposal generation. Frequentist method uses \texttt{scipy.optimize} to minimize the non-stationary negative log likelihood, while the Bayesian MCMC methods are implemented from scratch in \texttt{numpy}, allowing full transparency and customization. The concepts of non-stationarity and MCMC techniques used in \texttt{nsEVDx} are based on the foundational texts by \cite{christian_p_robert_introducing_2009, coles_introduction_2007}. The implementation of L-moments in some utility methods follows the approach described by \cite{j_r_m_hosking_regional_1997}. Currently, \texttt{nsEVDx} supports linear modeling for the location and shape parameters, and exponential (log-linear) modeling for the scale parameter, to ensure positivity.

Non-stationarity is controlled via a configuration vector \texttt{config = [a, b, c]}, where each entry specifies the number of covariates used to model the location, scale, and shape parameters of the EVD. Entry with a value of 0 implies stationarity (i.e., no covariate dependence), while integer values $>0$ indicate non-stationary modeling using the corresponding number of covariates for the parameter.

In Bayesian estimation, \texttt{nsEVDx} can infer prior specifications based on the data and configuration or accept user-defined priors. In the frequentist mode, it can determine suitable parameter bounds automatically. However, user-defined priors or bounds are recommended for better convergence and interpretability.

Future updates will potentially include mixed population models using categorical covariates to represent different distributions, an emerging area in hydroclimatic extremes. Additionally, efforts to optimize and accelerate code execution for faster runtimes are planned.

\section{Installation}

Install the package via pip:

\begin{lstlisting}
pip install nsEVDx
\end{lstlisting}

or alternatively, clone the repository and install manually:

\begin{lstlisting}
git clone https://github.com/Nischalcs50/nsEVDx.git
cd nsEVDx
pip install .
\end{lstlisting}

\section{Example Usage}

\begin{lstlisting}
from nsEVDx import NonStationaryEVD
from scipy.stats import genextreme

sampler = NonStationaryEVD(data, 
                          covariate, config=[1,0,0], 
                          dist=genextreme)
# config = [1,0,0] means, location parameter is modeled linearly
# with covariate, while scale and shape are treated as stationary
# Priors are inferred from the data if not provided while 
# declaring the sampler

print(sampler.descriptions) # provides the parameter descriptions 

samples, acceptance_rate = sampler.MH_RandWalk(
    num_samples=10000,
    initial_params=[10, 0.02 , 5, 0.1], 
    # B0(location intercept), B1 (location slope), scale, shape
    proposal_widths=[0.01, 0.001, 0.01, 0.001],
    T=1.0
)
\end{lstlisting}

See full documentation at: \url{https://github.com/Nischalcs50/nsEVDx/docs/API.md}

\section{Acknowledgements}

I gratefully acknowledge the support and encouragement of my wife, Koshika Timsina, whose constant belief in me has been a source of strength throughout this project. I also extend my heartfelt thanks to my family for their unwavering love, patience, and support.

\bibliographystyle{unsrtnat}

\begin{thebibliography}{14}
\providecommand{\natexlab}[1]{#1}
\providecommand{\url}[1]{\texttt{#1}}
\expandafter\ifx\csname urlstyle\endcsname\relax
  \providecommand{\doi}[1]{doi: #1}\else
  \providecommand{\doi}{doi: \begingroup \urlstyle{rm}\Url}\fi

\bibitem[{Heffernan J. E.} et~al.(2003){Heffernan J. E.}, {Stephenson A.G.}, and {Gilleland E.}]{heffernan_j_e_ismev_2003}
{Heffernan J. E.}, {Stephenson A.G.}, and {Gilleland E.}
\newblock ismev: {An} {Introduction} to {Statistical} {Modeling} of {Extreme} {Values}, April 2003.
\newblock URL \url{https://CRAN.R-project.org/package=ismev}.
\newblock Institution: Comprehensive R Archive Network Pages: 1.42.

\bibitem[Gilleland(2025)]{gilleland_extremes_2025}
Eric Gilleland.
\newblock {extRemes}: {Extreme} {Value} {Analysis}, May 2025.
\newblock URL \url{https://CRAN.R-project.org/package=extRemes}.
\newblock Institution: Comprehensive R Archive Network Pages: 2.2-1.

\bibitem[Paciorek(2016)]{paciorek_climextremes_2016}
Christopher Paciorek.
\newblock {climextRemes}: {Tools} for {Analyzing} {Climate} {Extremes}, August 2016.
\newblock URL \url{https://CRAN.R-project.org/package=climextRemes}.
\newblock Institution: Comprehensive R Archive Network Pages: 0.3.1.

\bibitem[{IRSN}(2024)]{irsn_nsgev_2024}
{IRSN}.
\newblock {NSGEV}: {Non}-{Stationary} {GEV} {Time} {Series}, 2024.
\newblock URL \url{https://github.com/IRSN/NSGEV/}.

\bibitem[{Oriol Abril-Pla} et~al.(2023){Oriol Abril-Pla}, {Virgile Andreani}, {C. Carroll}, {L. Y. Dong}, {Christopher Fonnesbeck}, {Maxim Kochurov}, {Ravin Kumar}, {Junpeng Lao}, {Christian C. Luhmann}, {Osvaldo A. Martin}, {Michael Osthege}, {Ricardo Vieira}, {Thomas V. Wiecki}, and {Robert Zinkov}]{oriol_abril-pla_pymc_2023}
{Oriol Abril-Pla}, {Virgile Andreani}, {C. Carroll}, {L. Y. Dong}, {Christopher Fonnesbeck}, {Maxim Kochurov}, {Ravin Kumar}, {Junpeng Lao}, {Christian C. Luhmann}, {Osvaldo A. Martin}, {Michael Osthege}, {Ricardo Vieira}, {Thomas V. Wiecki}, and {Robert Zinkov}.
\newblock {PyMC}: a modern, and comprehensive probabilistic programming framework in {Python}.
\newblock \emph{PeerJ Computer Science}, 9:\penalty0 e1516--e1516, September 2023.
\newblock \doi{10.7717/peerj-cs.1516}.
\newblock MAG ID: 4386373863 S2ID: 27127b9617c4a66f2a4a89f7ca57695623b3c218.

\bibitem[development Team(2023{\natexlab{a}})]{noauthor_pystan_2023}
Stan development Team.
\newblock {PyStan}: {The} python interface to {Stan}, 2023{\natexlab{a}}.
\newblock URL \url{https://pystan.readthedocs.io/}.

\bibitem[development Team(2023{\natexlab{b}})]{noauthor_cmdstan_2023}
Stan development Team.
\newblock {CmdStan}: {The} command-line interface to {Stan}, 2023{\natexlab{b}}.
\newblock URL \url{https://mc-stan.org/users/interfaces/cmdstan}.

\bibitem[Roberts and Tweedie(1996)]{roberts_exponential_1996}
Gareth~O. Roberts and Richard~L. Tweedie.
\newblock Exponential {Convergence} of {Langevin} {Distributions} and {Their} {Discrete} {Approximations}.
\newblock \emph{Bernoulli}, 2\penalty0 (4):\penalty0 341, December 1996.
\newblock ISSN 13507265.
\newblock \doi{10.2307/3318418}.
\newblock URL \url{https://www.jstor.org/stable/3318418?origin=crossref}.

\bibitem[Betancourt(2017)]{michael_betancourt_conceptual_2017}
Michael Betancourt.
\newblock A {Conceptual} {Introduction} to {Hamiltonian} {Monte} {Carlo}.
\newblock \emph{arXiv: Methodology}, January 2017.
\newblock ARXIV\_ID: 1701.02434 MAG ID: 2572702041 S2ID: 59b3ebd49ef620b3051d9777dfc59076d846949f.

\bibitem[Kafle and Meier(2025)]{kafle_evaluating_2025}
Nischal Kafle and Claudio Meier.
\newblock Evaluating {Methodologies} for {Detecting} {Trends} in {Short}-{Duration} {Extreme} {Rainfall} in the {Southeastern} {United} {States}.
\newblock In \emph{Extreme {Hydrological} or {Critical} {Event} {Analysis}-{III}, EWRI Congress 2025, Anchorage, AK, {U.S.}}, Anchorge, AL, 2025. ASCE.
\newblock URL \url{https://alaska2025.eventscribe.net}.

\bibitem[Kafle and Meier(in preparation)]{kafle_detecting_nodate}
Nischal Kafle and Claudio Meier.
\newblock Detecting trends in short duration extreme precipitation over {SEUS} using neighborhood based method.
\newblock \emph{Manuscript in preparation}, in preparation.

\bibitem[Robert and Casella(2009)]{christian_p_robert_introducing_2009}
Christian~P. Robert and George Casella.
\newblock \emph{Introducing {Monte} {Carlo} {Methods} with {R}}.
\newblock November 2009.
\newblock \doi{10.1007/978-1-4419-1576-4}.
\newblock MAG ID: 2162188693 S2ID: 1cf894b52919a330c3b477024b203d03058b34e3.

\bibitem[Coles(2007)]{coles_introduction_2007}
Stuart Coles.
\newblock \emph{An introduction to statistical modeling of extreme values}.
\newblock Springer series in statistics. Springer, London Berlin Heidelberg, 4th. printing edition, 2007.
\newblock ISBN 978-1-85233-459-8.
\newblock URL \url{https://doi.org/10.1007/978-1-4471-3675-0}.

\bibitem[Hosking and Wallis(1997)]{j_r_m_hosking_regional_1997}
Jonathan R.~M. Hosking and James~R. Wallis.
\newblock \emph{Regional {Frequency} {Analysis}: {An} {Approach} {Based} on {L}-{Moments}}, volume~93.
\newblock Cambridge University Press, April 1997.
\newblock URL \url{https://ui.adsabs.harvard.edu/abs/1997rfa..book.....H}.
\newblock MAG ID: 2056199992 S2ID: dafa032edadcea535f84f60334af40a3ef1cef6c.

\end{thebibliography}

\end{document}